\documentclass[aps,prx,reprint,superscriptaddress,nobibnotes,longbibliography]{revtex4-2} 
\pdfoutput=1

\usepackage{graphicx}
\usepackage{amsmath,amssymb}
\usepackage{xspace}
\usepackage{siunitx}
\usepackage[colorlinks]{hyperref}
\usepackage[caption=false]{subfig}
\usepackage{textcomp}
\usepackage{xcolor}
\usepackage{soul} %strikethough command package
\usepackage{physics}

\begin{document}

\newcommand{\change}[1]{\textcolor{red}{#1}}
\newcommand{\avg}[1]{\ensuremath{\langle #1 \rangle}\xspace}%

\title{Quantum gas microscopy of a geometrically frustrated Hubbard system}

\author{Jirayu Mongkolkiattichai}
\thanks{Contributed equally}
\author{Liyu Liu}
\thanks{Contributed equally}
\author{Davis Garwood}
\author{Jin Yang}
\email[Corresponding author: ]{dypole\_jin@mit.edu; Present address: Department of Physics, Research Laboratory of Electronics, MIT-Harvard Center for Ultracold Atoms, Massachusetts Institute of Technology, Cambridge, Massachusetts 02139, USA}
\author{Peter Schauss}
\email[Corresponding author: ]{ps@virginia.edu}

\affiliation{Department of Physics, University of Virginia, Charlottesville, Virginia 22904, USA}
%\email{ps@virginia.edu}

\begin{abstract}
Geometrically frustrated many-particle quantum systems are notoriously hard to study numerically but are of profound interest because of their unusual properties and emergent phenomena. In these systems energetic constraints cannot be minimized simultaneously, leading to large ground-state degeneracy and a variety of exotic quantum phases.
Here, we present a platform that enables unprecedentedly detailed experimental exploration of geometrically frustrated electronic systems on lattices with triangular geometry. We demonstrate the first realization of triangular atomic Hubbard systems, directly image Mott insulators in the triangular geometry with single-atom and single-site resolution, and measure antiferromagnetic spin-spin correlations for all nearest neighbors allowing for thermometry. This platform provides a powerful new approach for studying exotic quantum magnetism and direct detection of quantum spin liquid signatures in Hubbard systems.
\end{abstract}

\maketitle

\section{Introduction}
Electronic systems typically establish long-range order at zero temperature. Surprisingly, there are systems that do not have this fundamental property. For example, quantum spin liquids \cite{Wannier1950,Anderson1987} form in the presence of conflicting energetic constraints that prevent long-range ordering. Interestingly, the absence of ordering opens the door to a variety of exotic phenomena. For example, quantum spin liquids can show fractional quasi-particle statistics analogous to those underlying the quantum Hall effect \cite{Wen1989}.

Time-reversal symmetry breaking has been predicted in numerical studies on frustrated systems and kinetic constraints caused by the frustration lead to complex time-evolution \cite{Balents2010,Batista2016,Zhou2017}. While frustrated systems with small number of particles can be accurately simulated with tremendous computational resources, predictions for the low-temperature phases in the thermodynamic limit are scarce and often debated \cite{Yoshioka2009,Shirakawa2017,Szasz2020}. Existing condensed matter realizations are complicated materials \cite{Balents2010}, making well-controlled model systems a valuable alternative for gaining insight into the physics of frustration. 

Ultracold atoms provide a unique way to explore quantum many-body physics through quantum simulations of frustrated quantum systems based on first principles. Prominent examples of quantum simulation with ultracold atoms include the realization of Hubbard models \cite{BlochDalibardZwerger2008} and the observation of many-body localization \cite{Gross2017}. While there is widespread evidence for insulating phases without magnetic ordering in frustrated Hubbard models, their existence and properties are still controversial on many lattice geometries, including the triangular lattice. Ultracold atoms in optical lattices implement Hubbard models \cite{Lewenstein2007,BlochDalibardZwerger2008,Esslinger2010}, where neighboring sites are coupled by hopping, and atoms on the same lattice site interact. Atomic Fermi-Hubbard systems were first realized with ultracold atoms in square lattices \cite{Joerdens2008,Schneider2008}. 
With the realization of quantum gas microscopes for fermions, it became possible to image fermionic atoms on the single-atom level \cite{Cheuk2015,Parsons2015,Haller2015,Edge2015,Omran2015}. Later, two-dimensional (2d) fermionic Mott insulators (MI) were detected with quantum gas microscopes using $^6$Li \cite{Greif2016} and $^{40}$K \cite{Cheuk2016a}. In particular, the characteristic antiferromagnetic correlations in the repulsive Hubbard model have been studied in detail \cite{Greif2013,Hart2015,Drewes2016b,Parsons2016,Boll2016,Cheuk2016,Brown2017}.

Here, we expand these capabilities to a triangular lattice structure as a paradigm for studies of geometric frustration \cite{Anderson1987}, and report on the site-resolved imaging of atomic Mott insulators in a triangular lattice. Geometric frustration does not preclude short-range correlations, and we measure these correlations to study Hubbard physics on the triangular lattice.

%figure 1
\begin{figure*}
\centering
\includegraphics[width=0.84\linewidth]{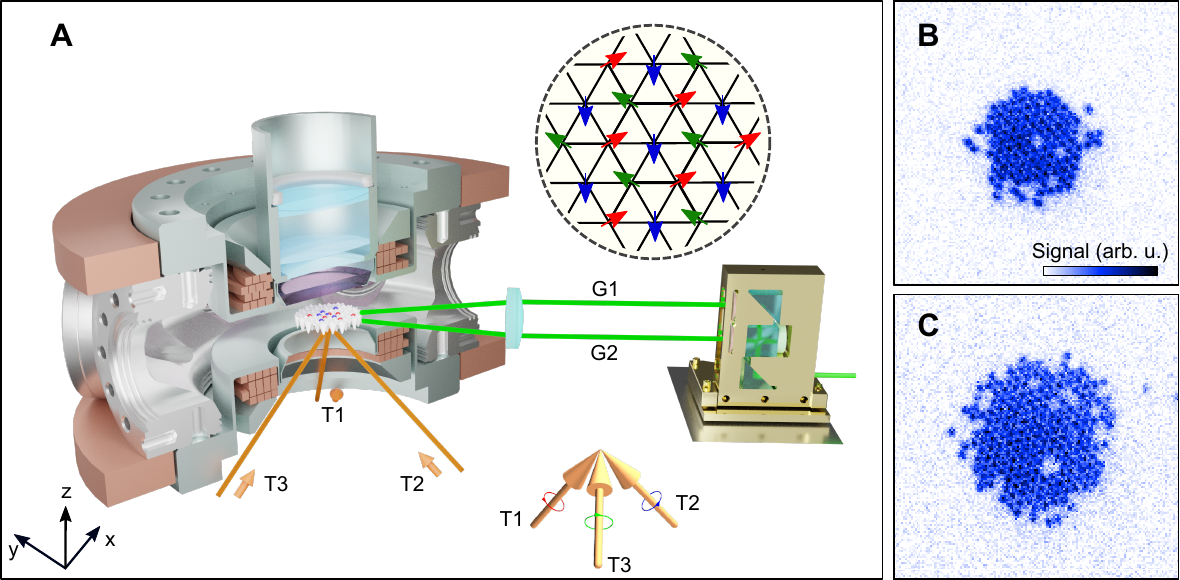}
\caption{{\bfseries Triangular-lattice quantum gas microscope.} {\bfseries{(A)}} A triangular optical lattice is realized by interfering three circularly polarized laser beams ($T1$, $T2$, and $T3$) using $1064$ nm light in the center of a vacuum chamber. The confinement of the atoms into two dimensions is achieved by a 1d accordion lattice in vertical direction, formed by $532$ nm laser beams. A combination of a beam splitter and mirrors allows us to vary the distance between $G1$ and $G2$ via the height of the input beam, therefore forming a lattice with a variable spacing between \SI{3}{\micro\meter} and \SI{8}{\micro\meter}. A high-resolution objective enables single-site resolved imaging of the atoms in the triangular lattice. The inset demonstrates $120^\circ$ order which is the classical analog of the spin ordering expected at large interactions. {\bfseries (B,C)} Triangular-lattice Mott insulators at $U/t=10(1)$ with $109$ atoms (top right) and $U/t=26(3)$ with $203$ atoms (bottom right). The field of view is \SI{40}{\micro\meter}$\cross$\SI{40}{\micro\meter}.}
\label{fig:schematic}
\end{figure*}

\section{Triangular-lattice Hubbard model}
The Hamiltonian of the fermionic system in a two-dimensional lattice at half-filling is 
\begin{equation}
    \begin{aligned}
    \mathcal{H} &= -t\sum_{\expval{\mathbf{r}\mathbf{r}\prime},\sigma} (c_{\mathbf{r},\sigma}^\dagger c_{\mathbf{r}\prime,\sigma} + c_{\mathbf{r}\prime,\sigma}^\dagger c_{\mathbf{r},\sigma}) + U \sum_{\mathbf{r}} n_{\mathbf{r},\uparrow} n_{\mathbf{r},\downarrow}\\ &-\mu(\mathbf{r})\sum_{\mathbf{r}} ( n_{\mathbf{r},\uparrow}+ n_{\mathbf{r},\downarrow})
    \end{aligned}
\label{eqn:Hubbard Hamiltonian}
\end{equation}
where $t$ is the tunneling strength between nearest-neighbor lattices, $U$ is the on-site interaction, $c_{\mathbf{r},\sigma}(c_{\mathbf{r}\prime,\sigma}^\dagger)$ is the annihilation (creation) operator for a fermion with spin $\sigma$ on site $\mathbf{r}$, $n_{\mathbf{r},\sigma}=c_{\mathbf{r},\sigma}^\dagger c_{\mathbf{r},\sigma}$ is the number operator, and $\mu(\mathbf{r})$ is the chemical potential.
This model describes the transition from a metal to a fermionic Mott Insulator, a prototypical example of a quantum phase transition. The insulating behavior originates from the electron-electron correlations and cannot be explained in a non-interacting electron picture. At temperatures below $U/k_B$, double occupation of sites is suppressed. Single occupation is energetically preferred at $\mu \sim U/2$ and the density variance vanishes, leading to a MI. When the chemical potential is larger than the energy gap, doublons (two atoms on a site) are formed. They first appear at the center of the trap because of the lower harmonic potential forming a band insulating core. More than two atoms per site are forbidden by the Pauli exclusion principle. Antiferromagnetic ordering can be observed in MIs when the temperature is comparable to the exchange energy 
$J={4t^2}/{U}$ \cite{Auerbach1990}. In the following, we present experimental data in this temperature regime and the observation of antiferromagnetic correlations on the triangular lattice.

\section{Experimental system}
We prepare a spin-balanced Fermi gas in a single layer of a one-dimensional (1d) accordion lattice (Fig.~\ref{fig:schematic}A) with a variable spacing. The gas is a mixture of the two lowest hyperfine ground states $\ket{\uparrow}=\ket{F=1/2,m_F=1/2}$ and $\ket{\downarrow}=\ket{F=1/2,m_F=-1/2}$ of $^{6}$Li, where $F$ and $m_F$ are the hyperfine quantum numbers (for details see \cite{SuppOnline}).
Next, the atoms are adiabatically loaded into the triangular lattice of depth $9.7(6) E_R$  using an s-shaped ramp. Here, $E_R=\hbar^2\pi^2/(2ma^2_\text{latt})=h\times8.2$ kHz is the recoil energy where $h$ is Planck's constant, $m$ is the atomic mass, and $a_\text{latt}=1003$ nm. The tunneling parameter is $t=h\times436(40)$ Hz \cite{SuppOnline}. The atom number and density in the lattice is adjustable by varying evaporation parameters. Once the atoms are in the lattice, we set the scattering length to $525(4)a_0$ where $a_0$ is the Bohr radius, thereby adjusting the interaction to $U/t=10(1)$. To detect the singles density ($n^s=n-n_{\uparrow} n_{\downarrow}$), the atom motion is frozen by linearly increasing the lattice depth to $100E_R$ within $8$ ms. For imaging, we turn off all magnetic fields and switch to maximum lattice depth. Images of MI for different interaction strengths are shown in Figs. \ref{fig:schematic}B and \ref{fig:schematic}C.

\begin{figure*}
\vspace{0cm}
    \centering
   \includegraphics[width=0.67\linewidth]{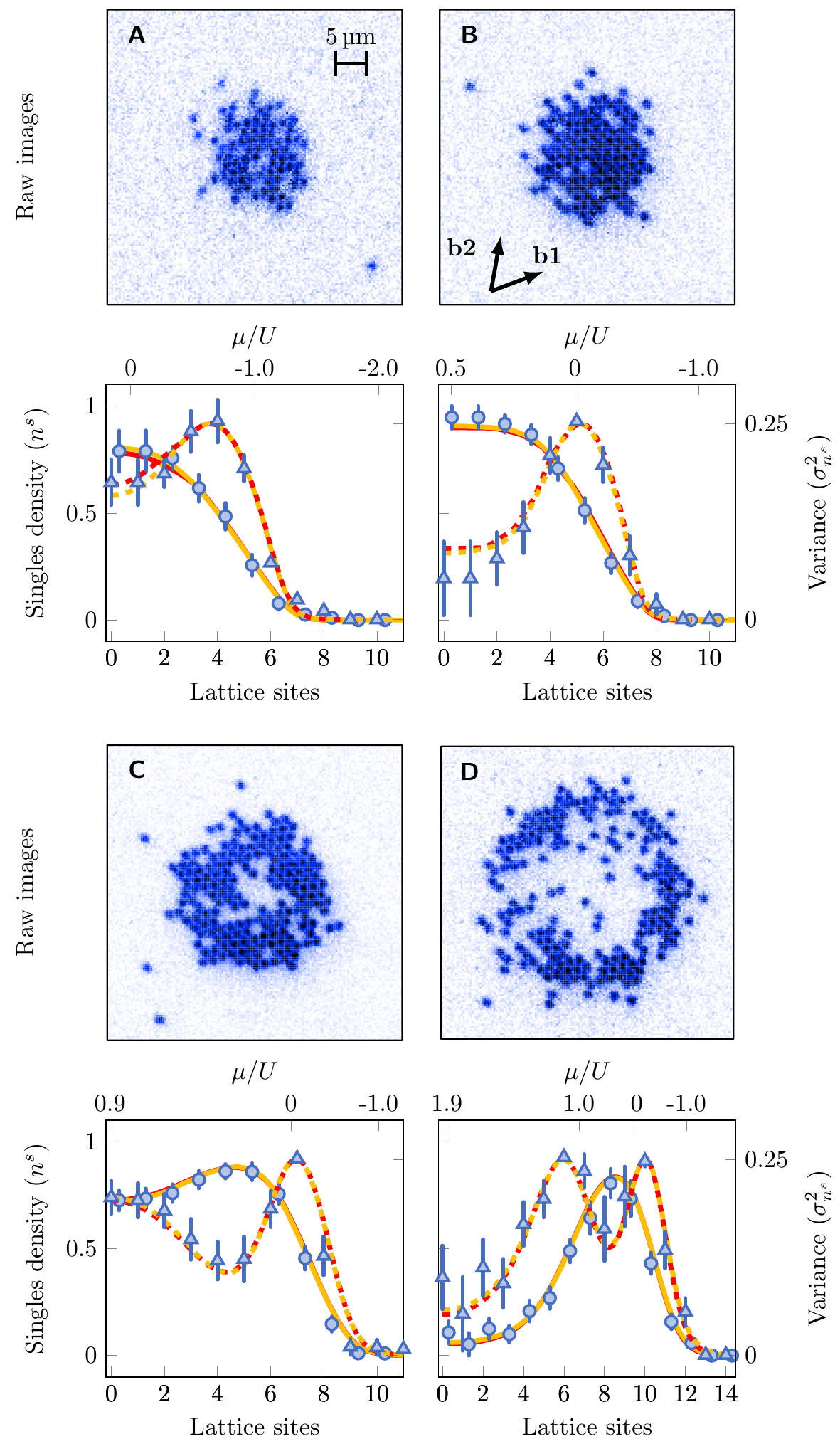}
\caption{{\bf Triangular-lattice Mott insulators.} {\bfseries (A-D) top} Site-resolved fluorescence images of fermionic Mott insulators with increasing atom number integrated from fit, $77$, $119$, $175$, and $183$ at interaction $U/t = 10(1)$. {\bfseries (A-D) bottom} Comparison of azimuthally averaged singles density (dots) and variance (triangles) with theory calculations, QMC (red) and NLCE (orange). The data points of the variance are horizontally offset by $0.3$ lattice sites for clarity. Both singles density $n^s$ and variance $\sigma^2_{n^s}$ are fit with QMC and NLCE theory using the local density approximation \cite{SuppOnline}. The detected variance is the square of the standard deviation of the sample within a radial bin. The fits yield temperatures $k_B T/t=0.9(2),\ 0.9(1),\ 1.5(1)$, and $2.4(1)$ with chemical potentials $\mu_0/U=0.24(10),\ 0.5(4),\ 0.91(3)$, and $1.94(1)$, respectively, at the trap center for increasing atom number in both QMC and NLCE calculations. Error bars on $n^s$ are standard error of the mean and error bars on $\sigma_{n^s}^2$ are determined by error propagation from $\sigma_{n^s}^2=n^s-(n^s)^2$.} 
\label{fig:mott}
\end{figure*}

By varying the atom number in the trap before loading atoms into the lattice, we observe MI and band insulators (BI) at $U/t=10(1)$ (Fig.~\ref{fig:mott}). The MI region (Fig. \ref{fig:mott}B) has nearly unit filling and atom number fluctuations are suppressed. When the chemical potential $\mu$ exceeds the value of $U/2$ (approximately half-filling), doubly occupied sites are formed, therefore a BI region in the center of the trap forms as shown in Figs. \ref{fig:mott}C and \ref{fig:mott}D. Doubly occupied sites are detected as empty sites due to light-induced collisions at the imaging stage \cite{Greif2016}.

%figure 3
\begin{figure*}
    \centering 
    \includegraphics[width=0.67\linewidth]{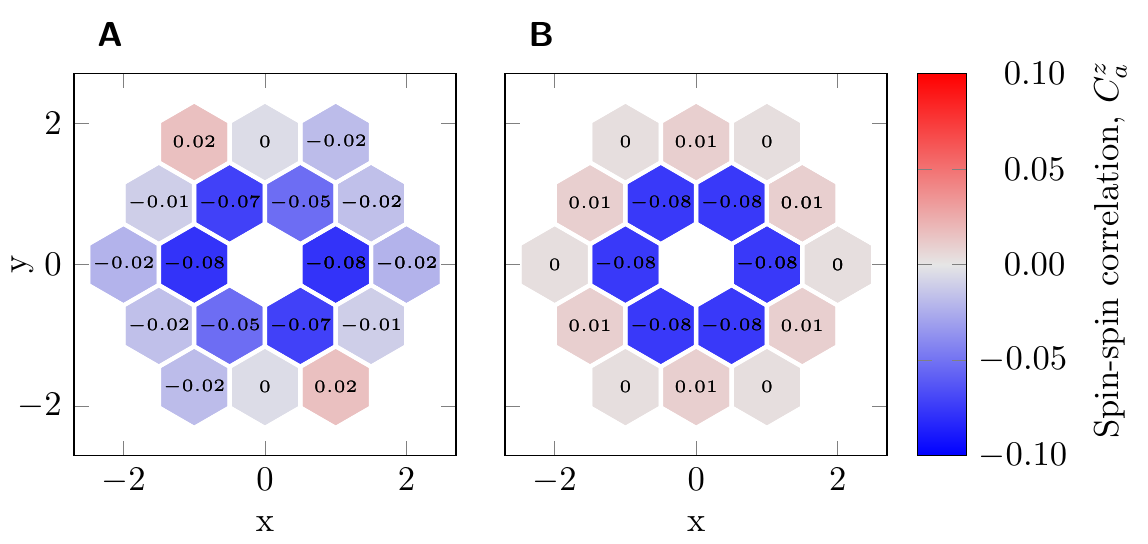}
    \caption{{\bfseries Spin-spin correlations}. {\bfseries(A)} Experimental correlations for $U/t=10(1)$. The $C^z_{\mathbf{b1}}$, $C^z_{\mathbf{b2}}$ and $C^z_{\mathbf{b1-b2}}$ are observed as anti-correlated along $(1,0)$, $(1/2,\sqrt{3}/2)$, and $(1/2,-\sqrt{3}/2)$. These values are the same within error bars suggesting tunneling isotropy of our triangular lattice. The correlations are extracted with post-selection from $400$ experimental pictures \cite{SuppOnline}. Typical values of experimental error bars are $\sim0.02$ and evaluated by bootstrap. {\bfseries(B)} Spin-spin correlations between nearest and next-nearest lattice sites calculated by DQMC at a temperature of $k_B T/t=0.8$ near half-filling and  $C^z_{0}$ is omitted for clarity. DQMC theory shows good agreement with experiment. The next-nearest-neighbor spin-spin correlations are consistent with zero.}
    \label{fig:spin_spin_correlations}
  \end{figure*}

\section{Triangular-lattice Mott insulators}
To access the singles density profile, we perform a deconvolution to determine the site occupation numbers and obtain singles density $(n^s)$ and variance $(\sigma_{n^s}^2)$ via azimuthal averaging (bottom panel of Fig.~\ref{fig:mott}). We fit the experimental density profile using determinantal quantum Monte Carlo (DQMC) and Numerical Linked Cluster Expansion (NLCE) calculations \cite{SuppOnline}. The temperature and chemical potential of the atoms in the trap are free parameters in the nonlinear least-squares fitting. We find good agreement with a global fit relying on a local density approximation using $\mu(\mathbf{r})=\mu_0-(1/2)m\omega^2r^2$ \cite{SuppOnline}. We obtain a temperature of $k_B T/t=0.9(1)$ and chemical potential of $\mu_0/U=0.50(4)$ at the trap center for the MI (Fig.~\ref{fig:mott}B) and a temperature of $k_B T/t=2.4(1)$ and chemical potential of $\mu_0/U=1.94(1)$ at the trap center for the BI (Fig.~\ref{fig:mott}D). We observe a small deviation at the center of the trap, which we attribute to the lower statistics and the uncertainty in the determination of the exact center of the system for azimuthal averaging. We observe an increased temperature for larger atom numbers as a result of reduced evaporative cooling.

\section{Spin-spin correlations}
Spin-spin correlations have proven to be essential observables for the understanding of the Hubbard model on square lattices \cite{Parsons2016,Boll2016,Cheuk2016,Brown2017}. Compared with the square lattice, we find that the magnitude of antiferromagnetic correlations in the triangular lattice is smaller, which we attribute to the geometric frustration.
The spin-spin correlator is defined as
\begin{equation}
    C_\mathbf{a}^z(\mathbf{r})=4\Big(\expval{S_\mathbf{r}^z S_\mathbf{r+a}^z}-\expval{S_\mathbf{r}^z}\expval{S_\mathbf{r+a}^z}\Big)
    \label{eqn:spin correlator}
\end{equation}
where the spin operator is  $S_\mathbf{r}^z=(n_{\mathbf{r},\uparrow}-n_{\mathbf{r},\downarrow})/2$. Here, the parameter $\mathbf{a}$ denotes the shift in the lattice site number between the two correlated positions, and $\mathbf{r}$ is the current lattice site. We access the observable $C_\mathbf{a}^z(\mathbf{r})$ via a linear combination of different correlators that can be measured directly in the experiment \cite{SuppOnline}. The fate of antiferromagnetic correlations on frustrated lattices is not obvious because the ordering is not compatible with the lattice structure. Despite the geometric frustration, we find significant antiferromagnetic correlations at nearest-neighbor sites. The reduced antiferromagnetic correlation, compared to the maximal correlation of $-1$ can be interpreted as incomplete anti-alignement of the spins. At large interactions, the Hubbard model maps to the Heisenberg model, and $120^\circ$ order is expected (Fig. \ref{fig:schematic}A). Negative nearest-neighbor correlations of $C_{\mathbf{b1}}^z=-0.078(22)$, $C_{\mathbf{b2}}^z=-0.053(23)$ and $C_{\mathbf{b1-b2}}^z=-0.071(28)$ are observed for three directions ($\mathbf{b1}$, $\mathbf{b2}$ and $\mathbf{b1-b2}$) as depicted in Fig.~\ref{fig:spin_spin_correlations}A \cite{SuppOnline}. We compare the experimental data with a correlation map calculated by DQMC at $U/t=10$ and $k_B T/t=0.8$ (Fig. \ref{fig:spin_spin_correlations}B). The calculated nearest-neighbor spin-spin correlations agree with the experimental data within error bars. The fact that all nearest-neighbor correlations are negative is consistent with $120^\circ$ order. Next-nearest-neighbor spin-spin correlations in the experimental data are consistent with zero within the typical uncertainty of ~$0.02$.

  %figure 4
 \begin{figure*}
    \centering
   \includegraphics[width=0.67\linewidth]{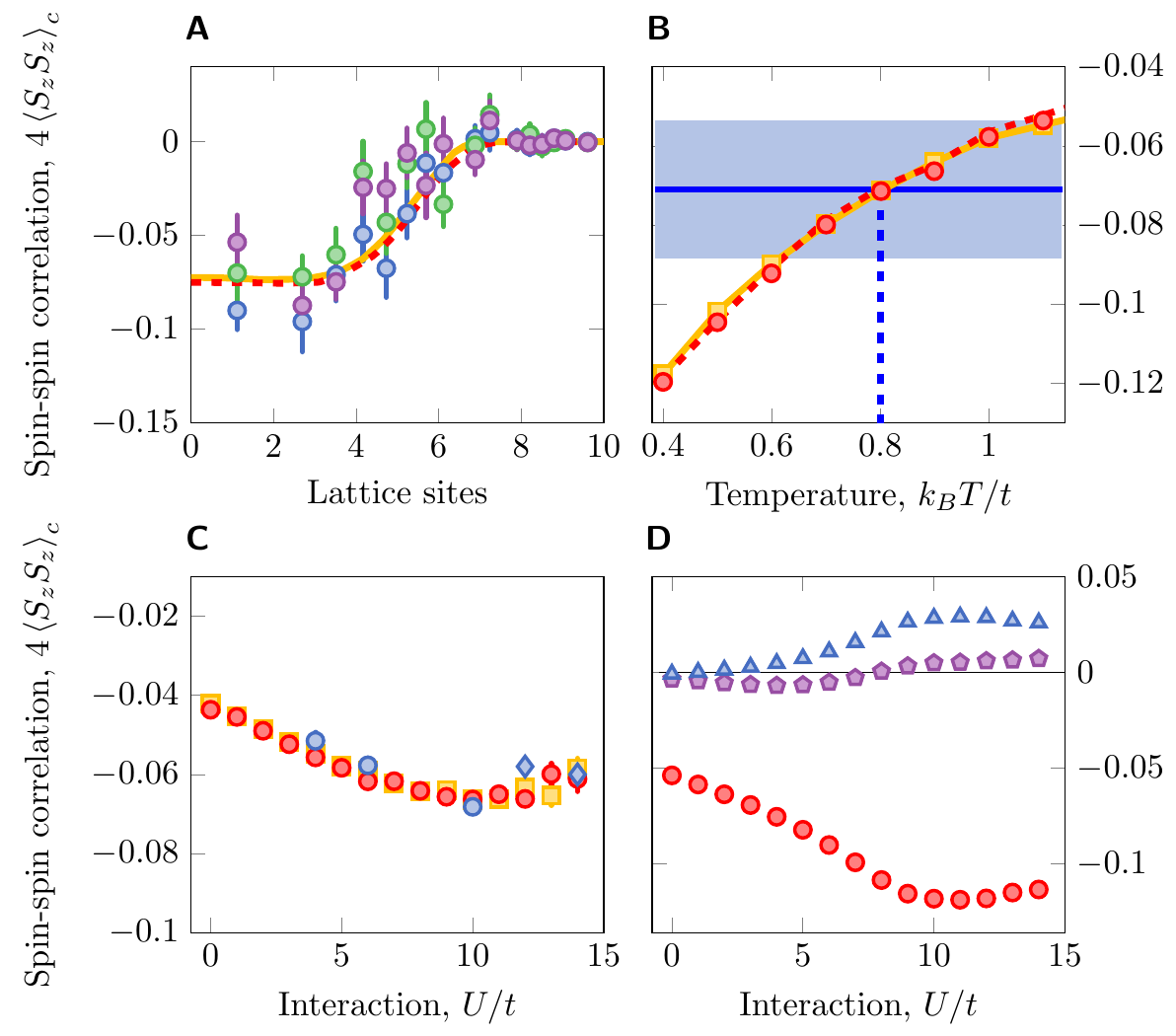}
\caption{{\bf Thermometry and interaction dependence of spin-spin correlations.} {\bfseries (A)} Spatial variation of correlations. Blue, violet and green dots are measured correlations along $\mathbf{b1}$, $\mathbf{b2}$ and $\mathbf{b1-b2}$, respectively. We perform an azimuthal average along the equipotential of the lattice confinement. The experimental data is fit to DQMC (red  dashed line) and NLCE (orange solid line) and extract a temperature $k_B T/t=0.80(10)$. Error bars are the standard error of the mean. {\bfseries (B)} Nearest-neighbor spin-spin correlation as a function of temperature. The experimental correlations at the center of the trap are visualized by the light blue shaded band with average indicated by the blue line compared with calculations from DQMC (red dots) and NLCE (orange squares) at half-filling. Orange solid and red dashed lines are a guide to the eye. The width of the blue band indicates the error of nearest-neighbor spin-spin correlation evaluated by the error propagation of nearest-neighbor spin-spin correlations established in Fig. \ref{fig:spin_spin_correlations}B. We find $k_B T/t=0.80(25)$ (blue dashed line). {\bfseries (C)} Interaction dependence of spin-spin correlations. Measured correlations (blue dots) are compared with DQMC (red dots) and NLCE (orange square) theory for temperature $k_B T/t\sim0.9$ at half-filling. Blue diamonds are measured using lattice depth of $12.0(7)E_R$ to avoid losses at larges values of $U$ \cite{SuppOnline}. Error bars are the standard error of the mean evaluated by bootstrap \cite{SuppOnline}. {\bfseries (D)} DQMC calculation of spin-spin correlations at $k_B T/t=0.4$ at $\mu = U/2$ (approximately half-filling) for shifts $(1,0)$, $(1.5,0.9)$ and $(2,0)$ (red dots, blue triangles and violet pentagons, respectively). The next-nearest-neighbor spin-spin correlations show a sign change versus $U/t$.}
\label{fig:SpatialCorr_and_TempDependence}
  \end{figure*}

\section{Thermometry}
To extract the temperature, we perform azimuthal averaging of nearest-neighbor correlations as a function of the distance from the trap center along the equipotential of the lattice confinement  \cite{SuppOnline} and fit to DQMC and NLCE calculations using temperature and chemical potential at the trap center as free parameters (Fig.~\ref{fig:SpatialCorr_and_TempDependence}A). We also average the correlations along the three lattice axes because they are equal within error bars. We show the result as a band in Fig.~\ref{fig:SpatialCorr_and_TempDependence}B and obtain a temperature of $k_B T/t\sim0.8$ by comparing correlations between experiment and theory calculations at half-filling. 
The measured temperature is consistent with the radial singles density fit in Fig.~\ref{fig:mott} with half-filling at the cloud center, providing evidence that the density and spin sector are thermalized. 

While this temperature is clearly below the interaction energy $U/t = 10(1)$, it is only slightly lower than the tunneling. We do not reach temperatures on the order of $k_B T/t\sim0.3$ which have been obtained in square lattices \cite{Mazurenko2017cold}. From comparisons to square lattice Mott insulators in our apparatus, we attribute the elevated temperature, in part, to the more complicated lattice geometry. We suspect that kinetic constraints due to the frustration make it harder to reach adiabaticity in the lattice loading. This issue deserves further theoretical and experimental study.

% U dependence discussion
In Fig.~\ref{fig:SpatialCorr_and_TempDependence}C, we show  nearest-neighbor spin-spin correlation versus interaction in comparison with DQMC and NLCE calculations for temperature $k_B T/t\sim0.9$ at half-filling. The strongest nearest-neighbor spin-spin correlations in the triangular lattice are found for $U/t \sim 10$ whereas the strongest correlations in the square lattice occur near $U/t \sim 8$ \cite{Cheuk2016}. We observe atom loss when increasing the scattering length beyond a value of $\sim650a_0$. Therefore, we change the lattice depth to reach larger $U/t$ \cite{SuppOnline}. 
We find good agreement with theory and note that experimental data lies on an isothermal graph for $k_B T/t\sim0.9$, indicating that the experimental temperature is almost independent of $U/t$.

% nearest, next-nearest spin-spin correlation vs interaction
Next-nearest-neighbor spin-spin correlations are challenging to measure as can be seen in Fig.~\ref{fig:SpatialCorr_and_TempDependence}D. DQMC calculations show a suppression of spin-spin correlations for next-nearest neighbors by a factor of $8$, compared to that for nearest neighbors, at a temperature $k_B T/t=0.4$ and half-filling. As interactions are increased, the next-nearest-neighbor spin-spin correlations are expected to cross over from negative to positive correlations in contrast to the situation in 2d square lattices at half-filling \cite{Parsons2016,Boll2016,Cheuk2016}. This will be the subject of future studies, as knowledge of the next-nearest neighbor correlations would make it possible to distinguish predictions for $120^\circ$ order and spin liquid correlations.

\section{Conclusion and Outlook}
In conclusion, we prepared atomic Mott insulators on a triangular optical lattice and performed single-site resolved imaging to detect spin-spin correlations. The radial density profiles of the observed Hubbard systems are in agreement with DQMC and NLCE calculations.
We find that thermometry based on nearest-neighbor correlations is possible in triangular Hubbard systems, despite the geometric frustration.

To observe longer-range correlations, temperatures in the experimental system will need to be lowered. Beyond the valuable information that could be gained from more extensive studies of heating within, and loading dynamics into, the frustrated lattice, entropy redistribution techniques are a promising path toward obtaining lower temperatures \cite{Mazurenko2017cold}.

Future experiments will access spin-density correlations in the system \cite{Boll2016,Koepsell2020}, enabling the study of polarons on the triangular lattice \cite{Vojta1999,Zhang2018,Kraats2022}. Binding energies are expected to scale with the tunneling $t$ and may be detectable at higher temperatures compared to square lattices \cite{Zhang2018}. Systems with increased binding energy are interesting because they may provide a path towards realizing repulsive pairing at higher temperatures and, therefore, higher-temperature superconductivity.
Additional future directions where our experimental platform can challenge state-of-the-art numerical calculations include the study of transport properties \cite{Vranic2020} and the experimental search for chiral ordering predicted for triangular Hubbard systems \cite{Wen1989,Szasz2020}.
\\
\\
\section*{Acknowledgments}
We thank Gia-Wei Chern, Bob Jones and Cass Sackett for careful reading of the manuscript. We thank Cass Sackett for sharing equipment.
We acknowledge support by the NSF (CAREER award \#2047275), the Thomas F. and Kate Miller Jeffress Memorial Trust and the Jefferson Trust.
D. G. was supported by a Ingrassia Scholarship.
J. M. acknowledges support by The Beitchman Award for Innovative Graduate Student Research in Physics in honor of Robert V. Coleman and Bascom S. Deaver, Jr.

\section*{Appendix}
\subsection{Preparation of the ultracold lattice gas}
\begin{figure}[h]
    \centering
    \includegraphics[width=0.9\linewidth]{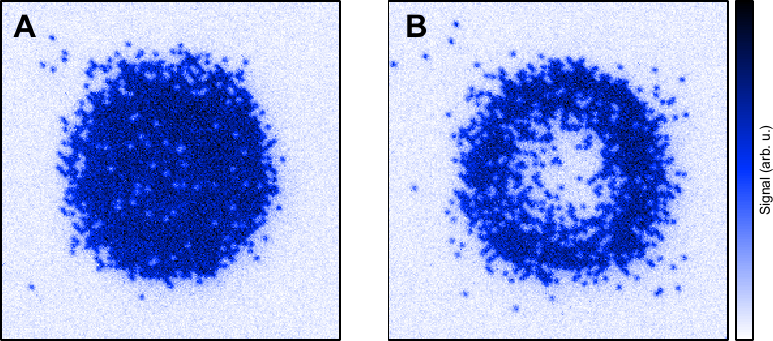}
    \caption{{\bfseries Square-lattice Mott and band insulators.} A $752$ nm square lattice is formed by four-beam interference using all vertical polarizations \cite{Brown2017,Sebby-Strabley2006}. {\bfseries (A)} Mott and {\bfseries (B)} band insulators consist of $1454$ and $1123$ atoms inside a radius of $21$ sites. The field of view is $60$ \textmu m$\times 60$ \textmu m.}
    \label{fig:Square Mott insulator}
\end{figure}

The procedure used to prepare a spin-balanced degenerate Fermi gas in a two-dimensional triangular optical lattice is described in detail in our earlier publication \cite{Yang2021}. We load the atoms directly from the magneto-optical trap (MOT) into a crossed optical dipole trap (ODT) and then load to a light sheet to create an oblate Fermi gas. Compared to the earlier work, we upgraded our coil-based Zeeman slower using permanent magnets and observe an improvement in loading rate of the magneto optical trap by a factor of $1.8(4)$ \cite{Garwood2022Hybrid}. From the light sheet, we load the atoms to a single layer of the accordion lattice, formed by $532$ nm laser beams, at largest spacing $8$ \textmu m in $z$ direction and provide horizontal confinement using a $1070$ nm laser beam (``bottom beam") with a Gaussian beam waist of $110$ \textmu m. The accordion lattice provides a variable vertical lattice spacing that facilitates a better loading efficiency. The accordion lattice is finally changed to $3$ \textmu m spacing, yielding a vertical trap frequency of $2\pi\times25(1)$ kHz. Next, we set the scattering length to $630(5) a_0$ \cite{Zuern2013} and perform evaporation in the accordion lattice by adjusting the bottom beam intensity in the presence of a magnetic gradient $\sim40$ G/cm along $z$ direction provided by the MOT coils. To study Hubbard physics, the symmetric triangular lattice is ramped up to $9.7(6) E_R$ and the Feshbach field is adjusted to obtain the target on-site interaction $U$. The triangular lattice formed by interference of three circularly polarized lattice beams is shown in Fig.~\ref{fig:schematic}A of the main text. We determine interaction and tunneling from lattice modulation and doublon formation spectroscopy, respectively. To freeze the motion of the atoms, the lattice depth is linearly increased to $100 E_R$ within $8$ ms. We then switch off all magnetic fields and increase the lattice depth to its maximum of $\sim~10^4 E_R$ for fluorescence imaging. The maximum lattice depth of this $\sigma$-configuration has been improved by a factor of three for the same lattice beam intensities and the same laser beam configuration as reported in ref.~\cite{Yang2021}. We obtain the sideband frequencies using Raman sideband spectroscopy. In addition to the triangular lattice, a square lattice is implemented where we observe the Mott insulator with a radius of $21$ lattice sites with $752$ nm lattice spacing (Fig. \ref{fig:Square Mott insulator}).

\subsection{Triangular lattice}
To form a triangular lattice, we cross three $\sigma$-polarized laser beams at $1064$ nm in contrast with the configuration reported in \cite{Yang2021}. This $\sigma$-polarization arrangement allows us to study Fermi-Hubbard model on a tunneling-isotropic triangular lattice.

An optical lattice potential is calculated by considering the electric field of lattice beams is given by
\begin{equation}
\begin{aligned}
\mathbf{E}_j(\mathbf{r})=&\sqrt{\frac{2I_j}{c\epsilon_0}}\exp\left[i(\mathbf{k}_j\cdot \mathbf{r}+\phi_j)\right]\times\\
&\left(\cos\theta_j\hat{\mathbf{e}}_{p,j}+\exp(i\alpha_j)\sin\theta_j\hat{\mathbf{e}}_{s,j}\right),\ j=1,2,3
\end{aligned}
\end{equation}
where $\sqrt{2I_j/c\epsilon_0}$ represents a field amplitude of lattice beam $j$ with intensity $I_j$ and $\hat{\mathbf{e}}_{p,j},\hat{\mathbf{e}}_{s,j}$ are unit vectors of $p$- and $s$- polarization of lattice beam $j$, respectively. $\theta_j$ determines the relative power between $s$- and $p$- polarization and $\alpha_j$ is the relative phase.

For our lattice configuration, the three lattice beams are tilted out of plane by $45^\circ$. This leads us to the following wavevectors for the three beams,
\begin{align}
    \mathbf{k}_1&=\frac{1}{\sqrt{2}}\begin{pmatrix}
1\\0\\1
\end{pmatrix}k_L,\\
\mathbf{k}_2&=\frac{1}{{2\sqrt{2}}}\begin{pmatrix}
-{1}\\{-\sqrt{3}}\\{{2}}
\end{pmatrix}k_L,\\
\mathbf{k}_3&=\frac{1}{{2\sqrt{2}}}\begin{pmatrix}
-{1}\\{\sqrt{3}}\\{{2}}
\end{pmatrix}k_L,
\end{align}
The magnitude of wavevectors is $k_L=2\pi/\lambda$ where $\lambda$ is laser wavelength. While each beam has $s$- polarization vectors
\begin{align}
    \hat{\mathbf{e}}_{s,1}&=\frac{1}{\sqrt{2}}\begin{pmatrix}
-1\\0\\1
\end{pmatrix},\\
\hat{\mathbf{e}}_{s,2}&=\frac{1}{{2\sqrt{2}}}\begin{pmatrix}
{1}\\{\sqrt{3}}\\{{2}}
\end{pmatrix},\\
\hat{\mathbf{e}}_{s,3}&=\frac{1}{{2\sqrt{2}}}\begin{pmatrix}
{1}\\{-\sqrt{3}}\\{{2}}
\end{pmatrix}.
\end{align}
and $p$- polarization vectors
\begin{align}
    \hat{\mathbf{e}}_{p,1}&=\begin{pmatrix}
0\\1\\0
\end{pmatrix},\\
\hat{\mathbf{e}}_{p,2}&=\frac{1}{{2}}\begin{pmatrix}
{\sqrt{3}}\\{-1}\\{0}
\end{pmatrix},\\
\hat{\mathbf{e}}_{p,3}&=\frac{1}{2}\begin{pmatrix}
{-\sqrt{3}}\\{-1}\\{0}
\end{pmatrix}.
\end{align}

By imposing $\alpha_j=-\pi/2$, $\theta_j=\pi/4$, and $\phi_j=0$, we calculate an optical lattice potential using $V(\mathbf{r})\propto \lvert\mathbf{E}_1+\mathbf{E}_2+\mathbf{E}_3\rvert^2$ and obtain
\begin{equation}
    \begin{aligned}
        V(\mathbf{r})=\frac{1}{4}V_0\Big[
        &\cos(\mathbf{b}_1\cdot\mathbf{r})+2\sqrt{6}\sin(\mathbf{b}_1\cdot\mathbf{r})\\
        +&\cos(\mathbf{b}_2\cdot\mathbf{r})+2\sqrt{6}\sin(\mathbf{b}_2\cdot\mathbf{r})\\
        +&\cos(\mathbf{b}_3\cdot\mathbf{r})-2\sqrt{6}\sin(\mathbf{b}_3\cdot\mathbf{r})
        \Big]
    \end{aligned}
\end{equation}
where $V_0$ is potential depth of a lattice beam, $\mathbf{b}_j$ denotes reciprocal vectors of lattice beams i.e., $\mathbf{b}_1=\mathbf{k}_1-\mathbf{k}_2$, $\mathbf{b}_2=\mathbf{k}_2-\mathbf{k}_3$, $\mathbf{b}_3=\mathbf{k}_1-\mathbf{k}_3$.

\begin{figure} [h]
    \centering
    \includegraphics[width=0.5\linewidth]{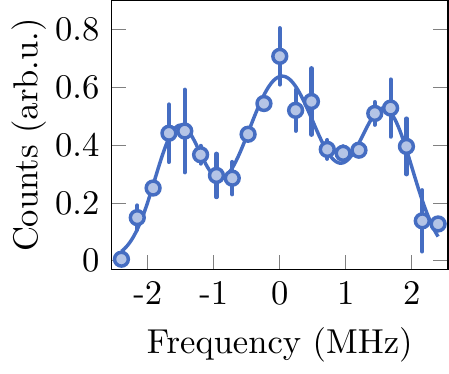}
    \caption{{\bfseries Raman sideband spectroscopy.} The
center peak is the bare transition between the $\ket{^2S_{1/2},F=1/2}$ and $\ket{^2S_{1/2},F=3/2}$ hyperfine states of the ground state whereas the two sidebands indicate the spacing of quantum harmonic oscillator in our optical lattice $\omega_\text{latt}=(2\pi)\times1.57(9)$ MHz. Dots represent experimental data and the solid line is a Gaussian fit to the three peaks. Error bars are the standard deviation of three repetitions.}
    \label{fig:raman_spectroscopy}
\end{figure}

We perform Raman sideband spectroscopy for the $\sigma$-polarization triangular lattice and obtain sidebands at $\omega_\text{latt}=(2\pi)\times1.57(9)$ MHz using the same beam intensity $\sim2.8\times10^6$ W/cm$^2$ and lattice configuration as described in \cite{Yang2021}. This is a major improvement in lattice depth without change in laser power. (Fig.~\ref{fig:raman_spectroscopy}).

\subsection{Atom loss at large scattering length}
\begin{figure} [h]
    \centering
    \includegraphics[width=0.5\linewidth]{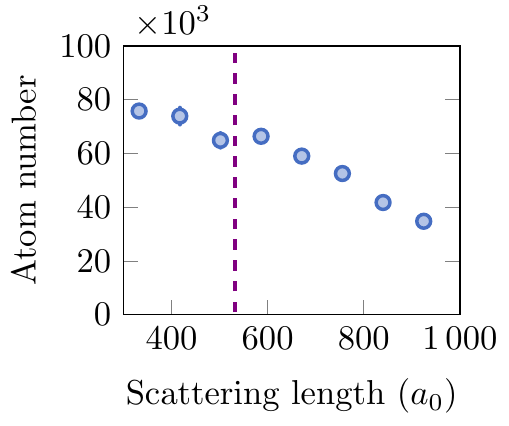}
    \caption{{\bfseries Atom number in the ODT as a function of scattering length.} Dots are experimental atom numbers and dashed line marks a position where the scattering length is $525a_0$. Error bars are the standard error of the mean.}
    \label{fig:atom number vs as}
\end{figure}

We measure atom number by holding atoms in the ODT for $4$ s and varying scattering lengths between $300a_0$ and $1000a_0$ (Fig.~\ref{fig:atom number vs as}). We observe more than $20\%$ atom loss for scattering lengths greater than $\sim650a_0$, therefore resulting in higher temperatures and weaker correlations for strong interactions at fixed lattice depth. Therefore, we limit the scattering length for most experiments to $525a_0$.

\subsection{Band structure calculation\label{sec.Band structure calculation}}
The band structure of triangular lattices is calculated using the plane-wave basis. We follow the approach in ref.~\cite{Flaschner2016}. The Hamiltonian of a non-interacting particle in an optical lattice potential is given by
\begin{equation}
    \hat{H}=\frac{\hat{p}^2}{2m}+\hat{V}
\end{equation}
The matrix elements in momentum space are given by
\begin{equation}
    H_{\mathbf{kk'}}\equiv\mel{\mathbf{k}}{\hat{H}}{\mathbf{k'}}=\int e^{i\mathbf{k}\cdot\mathbf{r}}\hat{H}e^{-i\mathbf{k}\cdot\mathbf{r}}\ d^3r
\end{equation}
where $\ket{\mathbf{k}}$ is a plane-wave basis of momentum $\mathbf{k}$ and it can be written in terms of reciprocal vectors $\mathbf{b_{1,2}}$ and quasi-momentum $\mathbf{q} $ i.e., $\mathbf{k}=n_1\mathbf{b_1}+n_2\mathbf{b_2}+\mathbf{q}$. Here, $n_1$ and $n_2$ are integer and $0<\abs{\mathbf{q}}<\abs{\mathbf{b_1}+\mathbf{b_2}}$.

\begin{figure} [h]
    \centering
    \includegraphics[width=\linewidth]{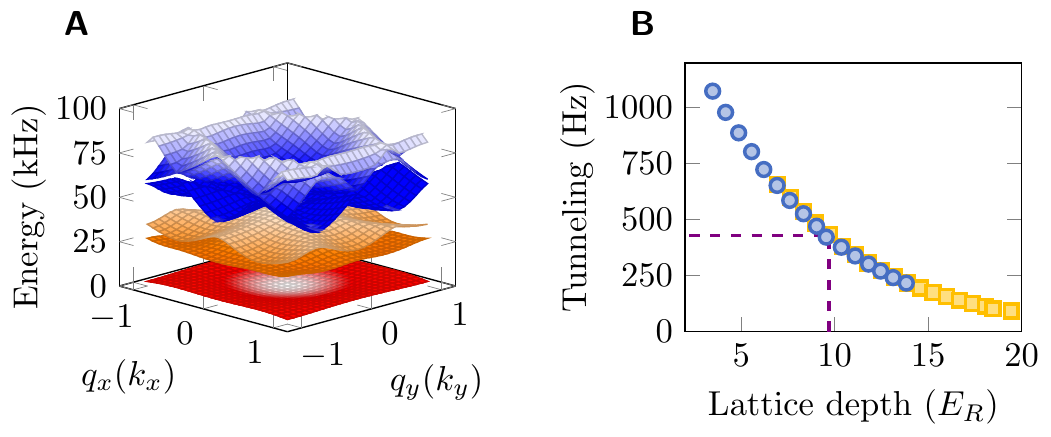}
    \caption{{\bfseries Triangular-lattice calculation.} {\bfseries (A)} Band structure calculated at a lattice depth of $9.7E_R$. {\bfseries (B)} Tunneling as a function of lattice depth using band structure calculation (blue dots) and Wannier functions (orange squares). The tunneling $t$ is 9 times the bandwidth of the $S$- band. The dashed line marks the lattice depth of $9.7E_R$.}
    \label{fig:BandStructure}
\end{figure}

Finally, eigenvalues of the Hamiltonian matrix depict the band structure shown in Fig. \ref{fig:BandStructure}A. We use the lowest band to calculate tunneling and the results agree with the method using Wannier functions (Fig. \ref{fig:BandStructure}B).

\subsection{Calculation of Hubbard parameters}
Wannier functions are the maximally localized wavefunctions on the sites of an optical lattice and can be used to calculate tunneling and interaction in tight binding approximation. To determine the Wannier functions, we initialize the Hamiltonian with kinetic energy, $\hat{T}$, and potential energy, $\hat{V}$, in the position basis. We then diagonalize the Hamiltonian and rewrite projection operators $\hat{\mathcal{P}}_X,\hat{\mathcal{P}}_Y$ in terms of the ground band of the Hamiltonian. We simultaneously diagonalize $\hat{\mathcal{P}}_X$ and $\hat{\mathcal{P}}_Y$. Next, we search for simultaneous eigenvectors of these projection operators \cite{bunse1993numerical}. Note that the determination of simultaneous eigenvectors has finite precision because our lattice is not exactly separable.
We finally transform the component in the lowest band to the position basis. This procedure is effectively the projection of a spatial delta function to the ground band which corresponds to the Wannier function on a site \cite{Kivelson1982}.

\subsubsection{Hubbard tunneling from Wannier functions}
The tunneling between two sites is determined by the overlap of two Wannier functions with the Hamiltonian $\hat{H}$ \cite{Blakie2004},
\begin{equation}
    t=\int w_0^\dagger(\mathbf{r})\hat{H}w_0(\mathbf{r-a})\ d^3\mathbf{r}.
\end{equation}
where $w_0$ is the Wannier function of the lowest band in the triangular optical lattice.

The tunneling is inferred from lattice depth calibrated using lattice modulation spectroscopy. We then calculate the tunneling parameter from the measured lattice depth.

\subsubsection{Hubbard interaction from Wannier functions}
On-site interaction $U$ can be obtained by integrating over two Wannier functions localized at the same lattice site whereas the Wannier function in vertical direction is treated as harmonic oscillator ground state wavefunction. Here, the on-site Hubbard interaction is given by
\begin{equation}
    U=\frac{4\pi\hbar^2a_0}{m}\sqrt{\frac{m\omega_z}{h}}\int|w_0(x,y)|^4\ dxdy
    \label{eq:on-site U formula}
\end{equation}
where $\omega_z$ is the confinement in the vertical direction.

\subsection{Calibration of Hubbard parameters}

\begin{figure} [h]
    \centering
    \includegraphics[width=0.5\linewidth]{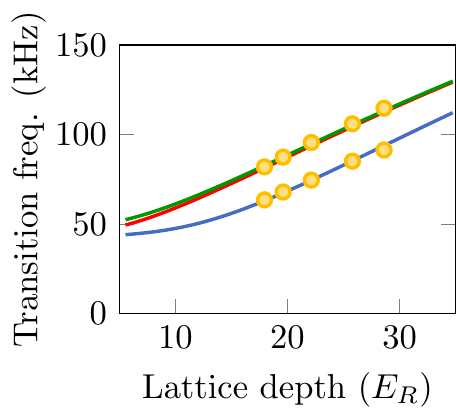}
    \caption{{\bfseries Band excitation spectroscopy.} Orange dots shows experimental data evaluated by Gaussian fit to atoms loss as a function of transition frequency. Solid lines are transition frequencies from $S$-band to $D$-band for different lattice depths calculated with band structure in the tight-binding limit. Transitions to the $P$-band are suppressed by the symmetry of the amplitude modulation. Error bars are smaller than the dots and evaluated by the standard error of the mean.}
    \label{fig:band_spectroscopy}
\end{figure}

\subsubsection{Band spectroscopy}
We calibrate our lattice depths in the range from $18E_R$ to $30E_R$ in the non-interacting regime at a Feshbach field of $527$ G. This configuration simplifies the Fermi-Hubbard model to a single-particle Hamiltonian approximated by a tight-binding model and we can calculate all band energies via a band structure calculation. We apply lattice modulation spectroscopy to characterize the lattice depth. After modulating the lattice beam power with an amplitude of $\sim 1\%$ for $20$ ms, we increase the lattice depth to maximum and measure atom number. We clearly observe two separated loss features whereas the third expected feature is overlapped with the second. The resonance is fit to our band structure calculations and we extract the lattice depth (Fig.~\ref{fig:band_spectroscopy}). The error bar of lattice depth is approximately $10$\% determined using a nonlinear fit to the band structure calculation with the lattice depth as a free parameter.

\subsubsection{Measurement of Hubbard interaction}

\begin{figure} [h]
    \centering
    \includegraphics[width=0.5\linewidth]{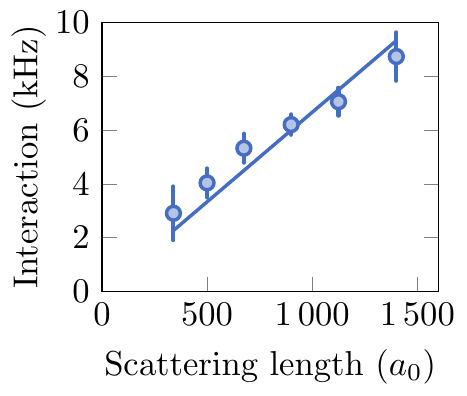}
    \caption{{\bfseries Doublon modulation spectroscopy.} Blue dots show detected interaction frequency as a function of scattering length. The interaction is determined by a decrease of singlons in fluorescence image due to doublon formation during modulation. We calibrate the magnetic field from the narrow Feshbach resonance at $543.3$ G by comparing the field to the scattering length from ref. \cite{Zuern2013}. The blue solid line is a linear fit to the data as expected in Eq. \ref{eq:on-site U formula}. Error bars are the standard error of the mean.}
    \label{fig:doublon_spectroscopy}
\end{figure}

To measure our Hubbard interactions, we prepare atoms in a lattice of depth $9E_R$ and perform amplitude modulation spectroscopy at varying Feshbach field corresponding to scattering lengths i.e., $500$, $675$, $900$, and $1400a_0$. We modulate the lattice with an amplitude of $\sim 5\%$ of the lattice depth for $20$ ms then measure atom number in fluorescence. When the modulation frequency is on resonance with the interaction energy, pairs of singlons form doublons \cite{Cheuk2016a} and we observe a decrease of ~$20$\% in detected atoms in fluorescence imaging due to light-assisted collisions \cite{DePue1999,Sherson2010}. According to Eq. \ref{eq:on-site U formula}, the on-site Hubbard interaction energy is linearly fit to scattering length ($a_0$) shown in Fig.~\ref{fig:doublon_spectroscopy} and we find the expected linear dependence within error bars.

\subsubsection{Tunneling isotropy}
We demonstrate that our lattice has isotropic tunneling within error bars by studying the spin-spin correlations which scale as $\sim\frac{t^2}{U}$ and are very sensitive to tunneling imbalance \cite{Brown2017}. 
We work at an interaction $U/t=10(1)$ where correlations are approximately maximal as predicted by DQMC and NLCE shown in Fig.~\ref{fig:SpatialCorr_and_TempDependence}C of the main text. We measure nearest-neighbor correlation functions along $\mathbf{b1}$, $\mathbf{b2}$, and $\mathbf{b1-b2}$ directions defined in the main text. The spatial dependence of the spin-spin correlations of all three nearest neighbors is shown in Fig.~\ref{fig:SpatialCorr_and_TempDependence}A of the main text. The three correlations are consistent within error bars, therefore we conclude that tunnelings in the triangular lattice are balanced within one standard deviation.

\subsection{Spin-spin correlations}
To extract the spin-spin correlation function $C_\mathbf{a}^z(\mathbf{r})$, we determine two types of correlations that can be directly calculated from datasets:  the single-species singles correlation for spin $\sigma$, $C_{\sigma}^{s}(\mathbf{r})$, and the singles correlation $C^{s}(\mathbf{r})$ \cite{Cheuk2016,Brown2017}. The spin correlator can be rewritten as
\begin{equation}
    C_\mathbf{a}^z(\mathbf{r})=2\Big(C_{\uparrow}^{s}(\mathbf{r})+C_{\downarrow}^{s}(\mathbf{r})\Big)-C^{s}(\mathbf{r})
    \label{eqn:exp correlator}
\end{equation}
where 
\begin{equation}
    C_{\sigma}^{s}(\mathbf{r})=\avg{n_{\sigma,\mathbf{r}}^{s}n_{\sigma,\mathbf{r+d}}^{s}}-\avg{n_{\sigma,\mathbf{r}}^{s}}\avg{n_{\sigma,\mathbf{r+d}}^{s}}
\end{equation}
and 
\begin{equation}
    C^{s}(\mathbf{r})=\avg{n_{\mathbf{r}}^{s}n_{\mathbf{r+d}}^s}-\avg{n_{\mathbf{r}}^{s}}\avg{n_{\mathbf{r+d}}^{s}}
\end{equation}
We obtain the single-species singles density ($n_{\sigma}^{s}$) by removing doubly occupied sites using doublon hiding \cite{Brown2017}  and remove either spin states using spin removal technique \cite{Parsons2016}. The singles density ($n^{s}$) is naturally measured during fluorescence imaging without any additional removal procedures.

To access the radial variation of correlations (Fig.~\ref{fig:SpatialCorr_and_TempDependence}A of the main text) we azimuthally average in lattice coordinates with a nearly equal chemical potential ($\lvert\Delta\mu/t\rvert \sim 0.6$ equivalent to $\Delta n^s\sim0.05$). This is necessary because the lattice confinement is strong for the lattice depth of $9.7(6)E_R$ and spin-spin correlations depend on the chemical potential. The error bars for the correlations are obtained by the standard deviation of correlation values in the equichemical bins divided by the square root of the number of correlation values. We average over the three symmetric neighbors which we justify by our demonstration of tunneling isotropy in our triangular lattice.

\subsection{Spin removal fidelity}
\begin{figure}[h]
    \centering
    \includegraphics[width=0.5\linewidth]{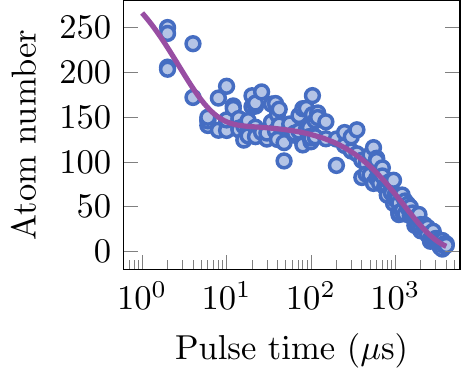}
    \caption{{\bfseries Atom number as a function of resonant light pulse time.} On-resonant pulse is applied for a variable time. Experimental data (blue dots) is fit to double decay function (violet solid line), resulting in $2.7(9)$ \textmu s and $1250(190)$ \textmu s for on-resonant and off-resonant removal time.}
    \label{fig:double decay}
\end{figure}

To image an individual spin component, we push out the other spin component and image the remaining atoms in fluorescence with a fidelity of $\sim98$ \% \cite{Yang2021}. To determine the imaging fidelity for this process, we prepare a MI at $U/t = 17$ then freeze the motion by ramping up the lattice depth to $100E_R$ within $8$ ms and reduce Feshbach field to the non-interacting point ($527$ G). We use vertical imaging beam with $I/I_{\text{sat}}=10$. The pulse duration is \SI{50}{\micro\second} determined by a double decay graph shown in Fig. \ref{fig:double decay} similar to ref.~\cite{Parsons2016}. By comparing the remaining atoms from the MI with and without pushing we can extract the spin removal fidelity to $\varepsilon_s=94(1)\%$ for both spins.

\subsection{Doublon hiding fidelity}
\begin{figure} [h]
\centering
\includegraphics[width=0.5\linewidth]{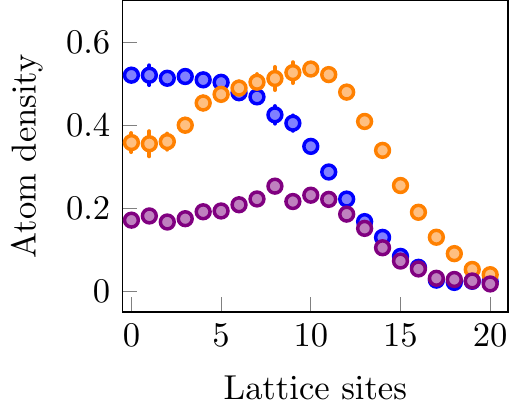}
\caption{ {\bf Doublon hiding fidelity.} To determine the doublon hiding fidelity, we need to take into account imperfections in the band insulator. The core region of band insulator contains a mix of single and double occupations ($n_{s}^{\text{BI}}$, orange) due to finite temperature. To determine an accurate fidelity nevertheless, doublon hiding and spin-removal techniques are applied to the band insulator ($n_{s,hp}^{\text{BI}}$, violet) whereas only spin-removal technique is performed to the band insulator resulting in the transformation from doublons to singlons ($n_{s,p}^{\text{BI}}$, blue). We apply Eqs. \ref{eq:formula1 for hiding fidelity}, \ref{eq:formula2 for hiding fidelity}, and \ref{eq:formula3 for hiding fidelity} to the core region within three lattice sites from center, therefore yielding the fidelity of doublon hiding technique $\eta_h=98(6)\%$.}
\label{fig:Doublon_hidding}
\end{figure}

We observe that during removal of one spin component, doublons are not reliably lost via light-induced collisions. Therefore, we apply a doublon removal technique \cite{Brown2017},  to eliminate doubly occupied sites before applying spin removal. After we ramp up our lattice to $100E_R$ we slowly sweep the Feshbach field over the narrow Feshbach resonance centered at $543.3$ G from high to low magnetic field. The fidelity of doublon hiding is determined by preparing a band insulator, and taking three separate datasets: no doublon hiding and no spin removal $(n_{s}^{\text{BI}})$, only spin removal ($n_{s,p}^{\text{BI}})$, and both doublon hiding and spin removal $(n_{s,hp}^{\text{BI}})$. 
Here, we obtain three singles densities from the measurements of a spin-balanced gas and consider a band-insulating core (Fig.~\ref{fig:Doublon_hidding}),
\begin{align}
    n_{s}^{\text{BI}}&=1-n_{d}^{\text{BI}},
    \label{eq:formula1 for hiding fidelity}
    \\
    n_{s,p}^{\text{BI}}&=n_{d}^{\text{BI}}\varepsilon_d+\frac{1}{2}\varepsilon_s n_{s}^{\text{BI}},
    \label{eq:formula2 for hiding fidelity}
    \\
    n_{s,hp}^{\text{BI}}&=n_{d}^{\text{BI}}\varepsilon_d(1-\eta_h)+\frac{1}{2}\varepsilon_s n_{s}^{\text{BI}},
    \label{eq:formula3 for hiding fidelity}
\end{align}
where $n_{d}$ is doublon density, $\varepsilon_d$ is spin-removal fidelity of doublons being singlons, $\varepsilon_s$ is spin-removal fidelity of singlons and $\eta_h$ is doublon hiding fidelity.
Note that we assume the majority at the center of the band insulator is doublons and singlons. Only doublons are lost during the doublon hiding because of the formation of weakly bound molecules. We solve these equations, resulting in the doublon hiding fidelity $\eta_h=98(6)\%$.

\subsection{Post-selection for spin-spin correlations}
We have technical atom number fluctuations in the experiment and therefore need to postselect to remove pictures with too low atom numbers from the dataset for spin-spin correlations. We choose pictures with a singles density larger than $0.84$ within a radius of $3$ lattice sites around the center. For pictures in which one spin component was removed (and doublon hiding was performed) we adjust the criterion to a density of $0.42$. This criterion is based on the average singles density of $\sim 0.86$ with the same lattice radius determined by DQMC theory for temperature $k_B T/t\sim1$ at half-filling. The post-selected dataset contains $\sim30$\% of all pictures equivalent to $\sim 400$ post-selected experimental pictures.

\subsection{Error estimates for spin-spin correlations}
For the statistical uncertainty of the correlators, we use a bootstrap technique ~\cite{bootstrap1986} which relies on a Monte Carlo algorithm. For different experiment conditions, e.g., interactions at $U/t = 10$, a typical dataset contains $\sim150$ pictures. Although the real probability distribution of the correlators requires infinite number of pictures, these $150$ pictures produce an empirical probability distribution. First of all, we pick samples from the dataset with an equal probability of $1/150$. Bootstrap sampling is expected to yield the same average but varying scatter as we randomly draw samples with size of $150$ with replacement from the real distribution. From each sample, we calculate values for the correlations. After repeating this procedure many times, the standard deviation of  calculated correlations approaches the one of the real probability distribution. The number of these bootstrap samples is typically in a range from $50$ to $200$. We sample until we reach convergence, which happens typically after around $100$ samples.

Systematic errors mainly arise from imperfections of spin-removal technique. We follow the procedure described in \cite{Cheuk2016} and obtain the error in the spin-spin correlator
\begin{equation}
    \begin{aligned}
        \Delta C^{z,\text{push}}_{\mathbf{a}}&\simeq\sum_\sigma\Big[-(\epsilon_{1\sigma}+\epsilon_{2\sigma})C_\mathbf{a}^z-(\epsilon_{1\sigma}-\epsilon_{2\sigma})C^s\\
        &-(\epsilon_{1\sigma}-\epsilon_{1-\sigma})(C^{s}_{\sigma}-C^{s}_{-\sigma})\Big]\\
        &= -0.12(1)C^z_{\mathbf{a}}+0.06(3)C^s
    \end{aligned}
\end{equation}
by omitting higher orders. Here, the summation of the unintended losses of spin-$\sigma$ atoms when performing spin-removal imaging $(\epsilon_{1\sigma})$ and the imperfect removal of $-\sigma$ atoms $(\epsilon_{2\sigma})$ is $6(1)\%$ based on our spin-removal fidelity. The term $\epsilon_{1\sigma}-\epsilon_{2\sigma}$ is $-3(2)\%$ determined using the double decay at the pulse time applied in the experiment. The last term is zero because we observe similar double decay for both spins.

The failure of doublon hiding could result in 
\begin{equation}
    \begin{aligned}
    \Delta C^{z,\text{hide}}_{\mathbf{a}}&\simeq\epsilon_d\avg{2\hat{n}_i\hat{d}_j+4\hat{d}_i\hat{d}_j}_c\\
    &=-0.0058(1-\eta_h)\\
    &=-1.16\times10^{-4}
    \end{aligned}
\end{equation}
where the infidelity of doublon hiding $\epsilon_d$ is $1-\eta_h$. We determine the density-doublon correlation $\avg{\hat{n}_i\hat{d}_j}_c$ using NLCE calculation for $U/t=10$ at half-filling and the doublon-doublon correlation $\avg{\hat{d}_i\hat{d}_j}_c$ is calculated by DQMC at the same interaction and filling.

The correction of the failure of the doublon hiding is negligible compared to the systematic error due to the spin-removal technique. Correlations presented in the main text take into account those corrections.

\subsection{Determinantal Quantum Monte Carlo calculations}
DQMC is an efficient technique to simulate the properties of fermions in lattices \cite{Blankenbecler1981,Paiva2010}.
We rely on a Fortran $90/95$ package, the QUantum Electron Simulation Toolbox (QUEST) \cite{Varney2009}.
Unfortunately, the calculations suffer from a severe sign problem for triangular Hubbard systems when approaching low temperatures \cite{Iglovikov2015} and we rely on extensive averaging for low temperatures. Below $k_B T/t \sim 0.4$ calculations are unreliable because the sign is approaching zero within error bars. Although the sign problem is severe, reliable results were obtained down to sufficiently low temperatures to allow for comparisons with the experiments.

Simulations rely on a homogeneous $8\times8$ lattice with periodic boundary conditions. We confirmed that for the properties discussed in this manuscript, finite-size errors are smaller than combined Trotter and statistical error. The inverse temperature $\beta=L d\tau$ was split into $L=40$ imaginary time slices for the data shown in Figs. 2, 3B, and 4A-C. To obtain higher statistics, the simulations were averaged over ten runs, $5,000$ warmup sweeps and $20,000$ measurement sweeps each. We use $L = 200$ time slices and 400 runs ($20,000$ passes each) for improved precision in Fig.~\ref{fig:SpatialCorr_and_TempDependence}D.  Multiple runs (with different seed) per Hubbard parameter set allow us to control imaginary time correlations and sampling errors by comparing the variance from each individual run with the variance of all runs.

\subsection{Numerical Linked Cluster Expansion calculations}
We use the NLCE code developed in our group described in detail in ref.~\cite{Garwood2022}. All NLCE calculations presented here were performed up to clusters of $9$ sites with $4$ cycles of Wynn resummation to improve convergence. In Fig.~\ref{fig:SpatialCorr_and_TempDependence}, nearest-neighbor spin-spin correlations are determined at $\mu = U/2$. For the triangular Hubbard model, the particle-hole symmetry is broken and the half-filling point deviates slightly from $U/2$ at low temperatures. For the temperatures shown, the deviation is much less than the size of the points in the plot \cite{Garwood2022}.

\subsection{Thermometry}
To determine the temperature $T$ of our Hubbard systems we apply two approaches: we fit the spatial variation of the spin-spin correlations over the system size and we average the spin-spin correlation at half-filling in the center of the system. Both techniques yield consistent results within error bars.
\subsubsection{Thermometry using density dependence of spin-spin correlations}
We fit radial nearest-neighbor correlation profiles to DQMC and NLCE (Fig.~\ref{fig:SpatialCorr_and_TempDependence}A). Specifically, we minimize the $\chi^2$ value defined by
\begin{equation}
    \chi^2(T,\mu)=\sum_{\mathbf{a}}\sum_{j}\Big(\frac{C_\mathbf{a}^{z}(\mathbf{r}_j)-C_\mathbf{a}^{z,\text{DQMC}}(\mathbf{r}_j)}{\Delta_{C_\mathbf{a}^{z}(\mathbf{r}_j)}}\Big)^2
\end{equation}
We take in account the symmetry of the lattice and the optimal fit therefore results in values for the temperature, $T$, and the chemical potential at the center of the trap, $\mu_0$.

In the fit, we calculate the correlations as a function of distance from the center of the system by employing a local density approximation (LDA),
\begin{equation}
    \mu(\mathbf{r})=\mu_0-\frac{1}{2}m\omega^2r^2
\end{equation}
where $m$ is atom mass and $\omega$ is harmonic lattice confinement.

We determine the lattice confinement by observing the best fit to the radial dependence of correlations at different lattice depths. At the lattice depth $9.7(6)E_R$ we obtain the lattice confinement of $2\pi\times540$ Hz. This value is consistent with calculations from the lattice beam parameters.

\subsubsection{Thermometry using spin-spin correlations at half-filling} 
We select correlations from the trap center within the radius of three lattice sites and average the values for all nearest neighbors. Then we extract temperature by comparing the averaged value to DQMC at half-filling as demonstrated in Fig.~\ref{fig:SpatialCorr_and_TempDependence}B of the main text.

\end{document}